\documentclass[12pt]{article}
\RequirePackage[OT1]{fontenc}
\usepackage{amsthm,amsmath,amssymb}
\usepackage{natbib}
 \bibpunct{(}{)}{;}{a}{,}{,}
\usepackage{stmaryrd}
\usepackage{graphicx}
\usepackage{hyphenat}
\usepackage{multirow}
\usepackage{multicol}
\usepackage{supertabular}
\usepackage{epsfig}
\usepackage{array}
\usepackage[figuresleft]{rotating} 
\usepackage{color}
\usepackage[vlined]{algorithm2e}
\usepackage[pdftex,colorlinks=true,linkcolor=blue,citecolor=blue,urlcolor=blue,bookmarks=false,pdfpagemode=None]{hyperref}
\usepackage{mathbbol}
\usepackage{setspace}

\numberwithin{equation}{section} 
\theoremstyle{plain}

\newcommand {\be} {\begin{equation}}
\newcommand {\ee} {\end{equation}}
\newcommand {\by} {\begin{eqnarray*}}
\newcommand {\ey} {\end{eqnarray*}}
\newcommand {\byn} {\begin{eqnarray}}
\newcommand {\eyn} {\end{eqnarray}}
\newcommand {\bd} {\begin{displaymath}}
\newcommand {\ed} {\end{displaymath}}

\usepackage{enumitem}

\usepackage[disable,colorinlistoftodos]{todonotes} 

\def\abovestrut#1{\rule[0in]{0in}{#1}\ignorespaces}
\def\belowstrut#1{\rule[-#1]{0in}{#1}\ignorespaces}

\newcommand{\para}[1]{\medskip\noindent{\it #1.}\quad}

\begin{document} 
\title{Estimating cellular pathways from an ensemble of heterogeneous data sources}
\author{Alexander M. Franks$^\dagger$, Florian Markowetz$^\ddagger$ and Edoardo Airoldi$^\dagger$\bigskip\\
$^\dagger$ Department of Statistics, Harvard University\medskip\\
$^\ddagger$ Cancer Research UK Cambridge Institute
}
{\makeatletter
\renewcommand*{\@makefnmark}{}
\footnotetext{This work was supported, in part, by NIH grant R01 GM-096193, NSF CAREER grant IIS-1149662, and by MURI award W911NF-11-1-0036 to Harvard University. EMA is an Alfred P. Sloan Research Fellow. Address correspondence to: \url{airoldi@fas.harvard.edu}.}
\makeatother}
\date{}
\maketitle

\thispagestyle{empty}
\newpage
\onehalfspacing

\begin{abstract}
Building better models of cellular pathways is one of the major
challenges of systems biology and functional genomics. There is a need
for methods to build on established expert knowledge and reconcile it
with results of high-throughput studies.  Moreover, the available data
sources are heterogeneous and need to be combined in a way
specific for the part of the pathway in which they are most
informative.  
Here, we present a compartment specific strategy to integrate edge,
node and path data for the refinement of a network hypothesis.
Specifically, we use a local-move Gibbs sampler for refining pathway
hypotheses from a compendium of heterogeneous data sources, including
novel methodology for integrating protein attributes.  We demonstrate
the utility of this approach in a case study of the pheromone response
MAPK pathway in the yeast S. cerevisiae.



\end{abstract}

%


\newpage
\setcounter{tocdepth}{3}


\tableofcontents

\listoftodos

\newpage
\setcounter{page}{1}

\section{Introduction}

Cellular mechanisms are driven by interactions between DNA, RNA, and proteins working together in cellular pathways. However, the current
knowledge of information flow in the cell is still very incomplete \citep{Kirouac2012}.
Even in well established signaling pathways studied for decades in model organisms, newer approaches can discover
novel components \citep{Mueller2005} or cross-talk with other pathways
\citep{McClean2007}. In cancer, finding pathways underlying disease development can lead to new drug targets \citep{Balbin2013}.
This makes the dissection of
cellular pathways one of the major challenges of systems biology and
functional genomics.

One of the main obstacles to utilize high-throughput data in
refining known pathway models is the gap between the relatively
unbiased and hypothesis-free nature of generating genome-scale
datasets and the need for very focused, hypothesis-driven research
to test biological models in small or medium scale experiments
\citep{Hibbs2008}. While researchers in computational biology
usually start with a collection of data and reconstruct pathways
from it, experimental biologists often start with a specific pathway
hypothesis in mind and try to reconcile it with the evidence from
high-throughput screens.

Here, we contribute to bridging this gap by introducing a
comprehensive data integration strategy to refine a given pathway
hypothesis. Our approach is characterized by three key features:  First, we start with a \emph{specific pathway model} and assess how
well it is supported in a collection of complementary data sets.
 These
data sets are heterogeneous and informative for distinct cellular locations.
Second,  we exploit this fact by introducing a \emph{compartment-specific}
probabilistic model, where data types are only used for
reconstructing the parts of a pathway they are informative about. Third, we  explicitly include \emph{node properties} in our model. This
allows us to use data like protein phosphorylation states or protein
domains, which have so far been under-utilized for pathway structure
learning \citep{Ryan2013}. 

In this paper we show that our modeling approach can
assist experimentalists in planning future studies by assessing
which parts of a biological model are not well supported by data,
and by proposing testable extensions and refinements of a given
pathway hypotheses. We demonstrate the power of our approach in a
case study in the yeast \emph{S. cerevisiae}.

\para{Related work}
Pathway reconstruction is a well established field in computational
biology \citep{Hyduke2010,Markowetz2007}. 
Several features distinguish our pathway refinement methodology from existing network reconstruction methods.

Comprehensive data integration strategies on large data collections
were shown to be very successful in predicting protein function and interactions
\citep{Guan2012,Llewellyn2008,Guan2008a,Myers2005}.
These methods are very helpful for describing the global landscape of protein
function, but offer less insight into individual molecular
mechanisms and pathways. 
Our approach differs from methods to
refine pathway hypotheses from expression profiles of down-stream
regulated genes \citep{Gat-Viks2007}, because we integrate heterogeneous data sources in a
compartment-specific way.

We also differ from previous research on de-novo pathway
reconstruction. 
These methods can be classified by how they use information about edges, paths and nodes in the pathway diagram for structure learning.
\begin{itemize}[leftmargin=5mm,itemsep=0mm,topsep=2mm,parsep=1mm]
\item
Most approaches incorporate evidence for individual \emph{edges} in
the pathway diagram using phenotypic profiles \citep{Mulder2012,Wang2012} or gene expression measurements  \citep{Li2013,Balbin2013,Schaefer2005a, Friedman2004, Segal2003}, sometimes supplemented by additional data sources like transcription factor
binding data \citep{Bernard2005,Werhli2007} or protein-protein
interactions \citep{Gitter2013,Nariai2004,Segal2003a}. Other studies
completely rely on protein-protein interactions to predict pathways
\citep{Mazza2013,Scott2006}.

\item
Cause-effect relationships indicating \emph{paths} from perturbed genes to observed effects are exploited in methods like SPINE \citep{Ourfali2007},
physical network models \citep{Yeang2005}, nested effects models
\citep{Wang2013, Markowetz2007a,Tresch2008,Froehlich2007,Froehlich2008a} and others \citep{Lo2012,Yip2010}, with applications
including DNA damage repair \citep{Workman2006} and cancer signalling \citep{Knapp2013,Stelniec-Klotz2012}.

\item
\emph{Node information}, i.e. features of individual proteins or genes, has been found useful for assigning proteins to
pathways \citep{Hahne2008,Froehlich2008} but has so far been under-utilized in reconstructing 
pathway structure \citep{Ryan2013}.
\end{itemize}

Our method differs from de-novo pathway reconstruction in that we
start with a hypothesis pathway and identify which hypothesized edges
are supported by the data.  We also differ from other methods which
evaluate formal one and two sample network hypothesis tests
\citep{Yates2013}.  Our goal is not to explicitly to determine whether
our initial hypothesis is ``correct''-- on the contrary we assume a
priori that any initial hypothesis can be further refined and improved
upon.  In the spirit of FDR, we provide a list of edge probabilities
that can assist experimentalists in their future studies.  We assess
which parts of an existing biological model are not well supported by
a data as well as suggesting new edges which are supported by the data
but which are not part of the original hypothesis.  Further, we are
the first to integrate data about \emph{edges} and \emph{paths} as
well as \emph{nodes} in the pathway diagram.

\todo{Alex, can you describe how we differ from \citep{Yates2013}}

\para{Overview}
We describe a compartment-specific probabilistic graphical model for posterior inference on cellular
pathways in \emph{section~\ref{sec:model}}, which can extend and refine a
given biological model and predict novel parts of the pathway graph.
Our model comprehensively integrates the three general types of data on edges, paths, and nodes.  We demonstrate the utility of our
methods in a case study in
\textit{S. Cerevisae}  (\emph{section~\ref{sec:case}}) by 
first exploring the information content in different data sources individually (\emph{section~\ref{sec:eda}})  and then
evaluating results of posterior draws using both full data and
leave-one-out data (\emph{section~\ref{sec:predict}}).


\section{Integating high-dimensional responses of a cellular pathway}
\label{sec:model}



Given a set of a gene products, i.e., putative pathway members, we infer an undirected
network model using a local-move Gibbs sampler.  The
pathway model, is defined in terms of N nodes and the edges
between these pairs of nodes, $(n,m)$.  The edges are encoded by a binary
random variable, $X_{nm}$.  The collection of edge-specific random
variables defines the adjacency matrix, \textbf{X}, of the pathway
model.\todo{What is in $\mathcal{M}$ except for \textbf{X}?
  Parameters? Compartment map? We should list these parts early on.}

\para{Parameter estimation and posterior inference} The adjacency matrix \textbf{X}
corresponding to the pathway model is latent since we cannot directly
observe the edges.  Thus, the primary goal of our analysis is to do
posterior inference on the adjacency matrix, \textbf{X}, from a
collection of M data sets, $Y_{1:M}$.  Although we treat \textbf{X} as
latent, we differ from de-novo pathway reconstruction by incorportaing 
an informative hypothesis pathway which we use to train the models
for data sets $Y_{1:M}$ (see Section \ref{sec:case}).

By Bayes rule, the posterior distribution on a pathway model,
\begin{equation}
\label{eq:posterior}
 P(X \mid Y_{1:M},\Theta) \propto
 P(X \mid \Theta) \cdot P(Y_{1:M} \mid \textbf{X},\Theta),
\end{equation}

\noindent is proportional to the prior distribution on the pathway with the
likelihood of the data.  Here, $\Theta$ is a collection of prior parameters
introduced below.

We use a local Gibbs sampling strategy to sample pathway models from
posterior distribution in Equation \ref{eq:posterior}.  The sampler
explores the space of pathway models by adding or removing edges in
turn, one at a time.  Specifically, the edge $X_{nm}$ between gene
products $(n,m)$ is sampled according to a Bernoulli distribution, with probability
of success

\begin{equation}
\label{eq:bern}
 P(X_{nm} \mid X_{(-nm)},Y_{1:M},\Theta),
\end{equation}

\noindent where $X_{(-nm)}$ represents the set of edges without $X_{nm}$.

\subsection{A compartment map defines context-specific data contributions}  

We use five complementary data types: physical binding of protein
pairs (including yeast-two hybrid, mass spectrometry, and
literature-curated data), transcription factor-DNA binding assays,
gene knockout data, gene co-expression data, and node information
(including protein domains and differential phosphorylation
arrays) 
Importantly, different data sets can be very informative in specific
cellular locations while completely uninformative in others.  Thus,
before we define the data likelihoods in
section~\ref{data-likelihoods}, it is essential to exploit this fact
in our model.  We translate expected compartment localization of a
pair of gene products $(n,m)$ into a binary importance vector $\vec
b_{nm}$, which drives the inference process by selecting the most
informative data types for the compartments involved.

To instantiate the notion that different data are informative in
different cellular locations, we introduce an additional modeling
element: the compartment map, which contains three conceptual pathway
compartments directly based on the organisation of the cell: First, the \textit{cell membrane}, where receptor proteins sense signals from outside the cell; second, the \textit{cytoplasm}, where protein cascades relay these signals to transcription factor proteins that enter the third compartment, the \textit{nucleus}, to regulate the activity of target genes.
The compartment map, $\mathcal{C}$, is a $5 \times 3$ binary matrix that associates the three pathway compartments with the five data types
to indicate which data type is informative about molecular
interactions in which compartments (see Table 1).

In particular, each data set is described by a pair $(Y_i,T_i)$, \todo{I got confused here. What is the relation between $\vec
b_{nm}$, $\mathcal{C}$, and $T_i$? From the description it sounded to me at
first like $b$ and  $T$ were rows/columns of R?}where $Y_i$ denotes
the collection of measurements, and $T_i$ is five-level factor that
denotes the specific data type used to collect the data and indexes
the relevant row of $\mathcal{C}$. We can now revise the form of the  conditional distributions in Equation \ref{eq:bern},
\begin{eqnarray}
\label{eq:new-bern}
 && P(X_{nm} \mid X_{(-nm)},Y_{1:M},T_{1:M},\mathcal{C},\Theta) = \\ \nonumber
 && ~~~ = \frac{\mathcal{L}( X_{nm}=1, X_{(-nm)}\mid Y_{1:M},\Theta
   )}{\mathcal{L}( X_{nm}=1, X_{(-nm)}| Y_{1:M},\Theta) + \mathcal{L}(X_{nm}=0, X_{(-nm)}| Y_{1:M},\Theta)}
\end{eqnarray}

Overloading notation, we let $\mathcal{C}_t(n,m)$ be an indicator
reflecting whether the protein pair $(n,m)$ is informative for data type
t, based on the compartment map and the localizations of proteins n and m. This leads to the following
likelihood specification:
\begin{eqnarray}
\label{eq:odds}
 && \mathcal{L}(X_{nm}, X_{(-nm)}\mid Y_{1:M},\Theta) \propto\\ \nonumber
 && ~~~ = \prod_{k}^M\left[ P(Y_k \mid X_{nm}, X_{(-nm)}, T_k=t, \Theta)^{
    \mathcal{C}_t(n,m)}\right.\\
&& \left. ~~~~~~~~~~\times P(Y_k \mid X_{(-nm)}, T_k=t, \Theta)^{1- \mathcal{C}_t(n,m)}\right]
\end{eqnarray}
where the role of the indicator is to discard data collections from
data types that are expected to carry little information about the
protein pair of interest, according to information in $\mathcal{C}$.
That is, for any pair $(n,m)$, $\mathcal{C}_t(n,m)=0$ implies data set
$Y_k$ is conditionally independent of $(n,m)$ given the rest of
the pathway.  In this case, the data in $Y_k$ has no effect on the
conditional posterior probability of $X_{nm}$.

\subsection{Modeling high-dimensional data for nodes, edges and paths}\label{data-likelihoods}

Data of different types need to be modeled
differently. We focus on modeling five main data types: protein
interaction data, protein-DNA binding data, gene co-expression data, gene perturbation data,
and node attribute data (differential phosphorylation and protein
domains).  Below, we describe the likelihood functions corresponding
to the main data types of interest.

\para{Likelihood for protein interaction data}
Here, we consider a single data set $Y_{N\times N}$ obtained with
data type $T$ aimed at measuring physical protein binding
events (PPI). We reduce the likelihood of the data, $Y$, to a
function the false positive and false negative rates, $\alpha$ and
$\beta$. Given the pathway, $X$, we evaluate
\begin{equation}
\label{eq:lik-ppi}
 \mathcal{L}_{ppi} (Y \mid X, \alpha,\beta)
 = \alpha^{S_{10}} (1-\alpha)^{S_{11}} \beta^{S_{01}} (1-\beta)^{S_{00}},
\end{equation}
where $S_{xy}$ counts the number of edges for which $X_{nm}=x$ and $Y_{nm}=y$. For instance, $S_{10}$ is the number of false positives.

\para{Likelihood for protein-DNA binding data}
Here, we consider a single data set $Y_{N\times K}$ obtained with
data type $T$ aimed at measuring transcription factor-DNA
binding events (TF). Rather than hybridization levels (for ChIP-chip) or peaks (for ChIP-seq), we model the
$p$-values corresponding to binding events, which makes our model independent of the technology used to detect the binding event. 
We develop a mixture
model for the $p$-values, directly. Given the pathway, $X$, we expect to see a small $p$-value for protein $n$ binding nucleotide
sequence $m$ whenever  the edge $X_{nm}$ is present. On the
contrary, the $p$-values are uniformly distributed under
the null hypothesis of no binding events, $X_{nm}=0$. We evaluate
\begin{eqnarray}
\label{eq:lik-tf}
 &&\mathcal{L}_{tf} (Y \mid X,\gamma)
 = \prod_{n,m} \bigm[ \text{Uniform }(Y_{nm}) \cdot\mathbb{1}(X_{nm}=0) \nonumber \\
 && ~~~ +  \text{Beta }(Y_{nm} \mid \gamma,1) \cdot\mathbb{1}(X_{nm}=1) \bigm],
\end{eqnarray}
where $0 < Y_{nm} <1$ ($p$-value), and $0 < \gamma < 1$.
See a related beta-uniform mixture model introduced by
\citet{Poun:Morr:2003} in the context of multiple testing for
differential expression.


\para{Likelihood for knock-out data} Here we consider a data set $Y_{M\times N}$, where $Y_{mn}$ is the
log-two-fold change in expression of gene n, when gene m is knocked
out.  Let $Z_{mn}$ be a binary variable representing the existence of a
directed path from gene n to gene m, \textit{through a transcription factor}.  While we consider the set of
undirected pathway models, we temporarily impute directionality using
the fact that the cellular signal should flow from the cytoplasm to the
nucleus.  We model the knockout data as a mixture of normals:

\begin{eqnarray}
\label{eq:lik-ko}
&& \mathcal{L}_{ko} (Y \mid X, \sigma_0, \sigma_1) =\\ \nonumber
 && ~~~ = \prod_{n,m} \text{Normal} \left(Y | 0, \sigma_1
 \right)\mathbb{1}[Z_{mn}]+\text{Normal} \left(Y | 0, \sigma_0\right)\mathbb{1}(1-Z_{mn})
\end{eqnarray}

\noindent The standard deviations for change in
expression are represented by $\sigma_0$ (when there is no path
between the knockout and a target) and $\sigma_1$
(there is a path). The assumption is that $\sigma_1 > \sigma_0$ since we
expect a larger change in expression of n for knockout m when n and m are connected in the pathway.


\para{Likelihood for gene co-expression data}
Here, we consider a single data set $Y_{N\times N}$ aimed at measuring gene expression. 
Rather than hybridization levels (for microarrays) or the number of reads (for mRNA sequencing), we model correlations among
the profiles of pairs of genes, which again makes our model independent of the details of the measurement technology. We develop a mixture model for the
correlations, directly. Given the pathway, $X$, we expect to see
correlation between the expression profiles of two genes whenever
they are co-regulated. Similarly to \citet{Schaefer2005}, we use a
mixture model for the distribution of the sample correlation
coefficient $\hat{\rho}=y$ of the form
\begin{eqnarray}
\label{eq:lik-exp}
 &&\mathcal{L}_{expr} (Y \mid X,\delta,\kappa)
 = \prod_{n<m} \bigm[ P_0 (Y_{nm}\mid \kappa)
 \cdot\mathbb{1}(X_{nm}=0) + \nonumber \\
 && ~~~ P_1(Y_{nm} \mid \delta,1) \cdot\mathbb{1}(X_{nm}=1) \bigm]
\end{eqnarray}
When $X_{nm}=0$, we expect the two gene profiles to be uncorrelated.
Differently from \citet{Schaefer2005}, however, we chose a
distribution that puts more emphasis on higher correlation if we see
an edge in the model, $X_{nm}=1$, using a one-parameter beta
distribution,
\begin{equation}
\label{eq:lik-exp-one}
 P_1(y|\delta) = \text{Beta } (y \mid \delta,1).
\end{equation}

\para{Likelihood for node attributes data}
Here, we consider a single data set $Y_{M\times N}$ that lists
node-specific attributes such as protein domains from PFAM \citep{Punta2012} and SMART
\citep{Schultz1998,Letunic2012} databases, and
differential phosphorylation data \citep{Gruhler2005}. 
We develop novel techniques to model
protein attributes. Specifically, we model the likelihood of an attribute
conditional on the given pathway \textbf{X}.  We term our models for
node attributes ``relation regression.''
For differential phosphorylation data, $Y_{N\times 1}$, 
\begin{eqnarray}
\label{eq:lik-rlm}
 && \mathcal{L}_{node} (Y \mid X, \lambda,\sigma) =\\ \nonumber
 && ~~~ = \prod_n \text{Normal} \left( Y_n \mid \lambda_0 + \lambda_1 \frac{\sum_{m\neq n} Y_m\mathbb{1}(X_{nm}=1)}{\sum_{m\neq n} \mathbb{1}(X_{nm}=1)}, \sigma^2 \right)
\end{eqnarray}

\noindent In other words, the differential phosphorolation, $Y_n$, is assumed to be
linearly related to the mean differential phosphorolation of the
neighbors of node n.  Similarly, for the protein domain data,
$D_{N\times K}$, we use an auto-logistic 
regression to model the data.  Specifically, for $D_{nk}$, a binary variable
indicating the presence of domain k in protein n,


\begin{equation}
\label{eq:lik-rlm}
\mathcal{L}_{node} (D \mid X, \lambda) = \prod_{nk} P_{nk}^{D_{nk}}(1-P_{nk})^{(1-D_{nk})}
\end{equation}

where $$P_{nk} = \text{logit}^{-1} \left( \lambda_0 +
\sum_{j}\lambda_j \mathbb{1}\left[\sum_{m\neq n}
  D_{mj}\mathbb{1}(X_{nm}=1) > 0 \right]\right)$$

\noindent Here, logit($P_{nk}$) is linearly related to the presence of domains
in neighboring genes.  In both the normal and logistic regression cases, we fit the
regression coefficients, $\vec \lambda$, using our initial pathway hypothesis.  In the
logistic model, we use a weakly-informative Cauchy
prior  for the coefficients
\citep[]{Gelman2008}.  This controls for any overfitting and separation
problems that may occur. 

\para{Prior distribution on the space of pathway models}
In this study our focus lies on assessing the extent to which the
data support a pathway model $X$. We choose a block model
prior $P(X)$ over binary matrices of size $N\times N$ with edge
density fixed by compartment.  In general,
any informative prior distribution on graphs could be used here to
encode biological knowledge \citep{Isci2013,Mukherjee2008a}.


\section{Analysis of the pheromone response pathway in \textit{S. cerevisiae}}
\label{sec:case}
To demonstrate the efficacy of our approach, we examine the pheromone
response MAPK pathway in the yeast \textit{S. cerevisiae}. 
It offers the opportunity to
combine a large collection of datasets with a solid understanding of
the pathway structure. The pheromone pathway is the subject of
intense research efforts in computational  biology
as well as experimental biology \citep{Hara2012,Scott2006,Kofahl2004} and shows cross-talk to other MAPK pathways \citep{Nagiec2012,McClean2007,Gat-Viks2007}.

\para{Initial pathway construction} \label{sec:hypothesis}
\begin{sidewaysfigure}[http]
\vspace{.2in}
\includegraphics[width=0.91\textheight]{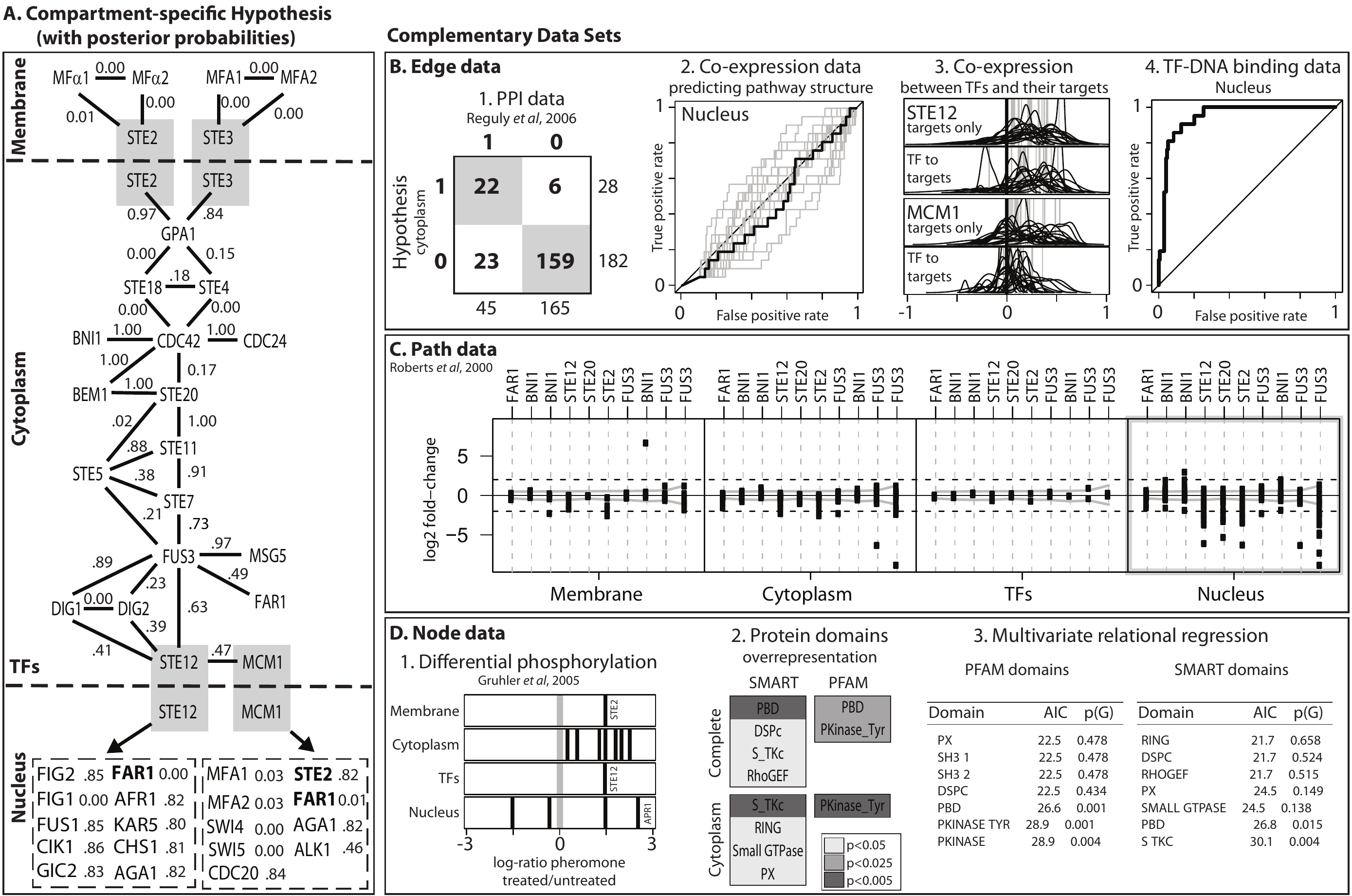}
\caption{\label{fig:data} Compartment-specific pathway hypothesis,
  posterior probabilities, and evaluation of support in the data. \textbf{A.} Pathway
hypothesis and posterior edge probabilities for the Yeast pheromone
response pathway. 
The numbers by each edge reflect the ``posterior
probability'' 
. 
\textbf{B.} Edge data: (1) protein-protein interactions in the
cytoplasm, (2) gene co-expression in the nucleus, (3) co-expression
of TFs with their targets and between targets for STE12 and MCM1,
and (4) TF binding data. \textbf{C.} Cause-effect data. 
\textbf{D.} Node data: (1) Differential phosphorylation,
(2) Overrepresentation of protein-domains in different compartments,
(3) goodness-of-fit of auto-logistic models on protein
domains from PFAM and SMART.}

\end{sidewaysfigure}
To start our analysis in a way relevant to refining and extending
existing knowledge of signaling pathways, we extracted a model of
the pheromone response pathway from the summary of MAPK pathways
(sce04010) in the database KEGG \citep{Kanehisa2000} and combined it
with known transcription factor (TF) targets from two independent
studies \citep{Simon2001,Ren2000}.

We split the pathway into three parts: the \emph{membrane}
compartment containing the receptor proteins, the \emph{cytoplasm}
compartment containing the MAPK cascade to activate the
transcription factors (TF), and the \emph{nuclear} compartment
containing the TFs and their targets. Figure~\ref{fig:data}-A
depicts the pathway hypothesis. Proteins mediating between two
compartments (like TFs) are contained in two sub-graphs and marked
by grey boxes. TF targets that are also members of other
compartments are indicated in bold.

\subsection{Exploratory data analysis of individual data types}
\label{sec:eda}

Before inferring the full model from all data, we explored the information content in each type of data individually.

\para{Protein-protein interactions (PPI)}
We compared data from several complementary high-throughput assays, all available
from BioGRID \citep{Stark2006}
as well as a
literature-curated dataset \citep{Reguly2006}. 
We analyzed the overlap between the protein interactions and the
pathway hypothesis of Fig~\ref{fig:data}-A. None of the datasets are
informative for the membrane and nuclear compartments. Surprisingly,
in the cytoplasm compartment we found that all of the
high-throughput datasets show  only $\leq3$ interactions between any
of the proteins in the pathway. 
The situation was very different
for the literature-curated data. Here, 45 interactions in the
cytoplasm compartment covered 22 out of the 28 edges there
(sensitivity $>78\%$, specificity $>87\%$, see
Fig~\ref{fig:data}-B1).

\para{TF-DNA binding data}
We used the transcription factor binding data of
\citep{Harbison2004}, which is independent of our definition of TF
targets in the pathways hypothesis. The ROC in
Figure~\ref{fig:data}-B4 shows very clear signal to distinguish the
targets posited in the biological model from all other pathway
genes.

\para{Co-expression data}
For gene expression data, we examined datasets in which the pathway
genes showed a significant difference in correlation structure from
all other yeast genes (using the SPELL algorithm of
\citep{Hibbs2007}) resulting in 20 datasets from 15 publications
\citep[including][]{Roberts2000,Gasch2000,Brem2005}. 
Figure~\ref{fig:data}-B2 shows ROCs for predicting edges in the nuclear compartment for all datasets (grey lines) and the concatenated data (black line).
No curve improves much on random prediction (the main diagonal).
%
The reason is biological: Because expression data are a poor surogate for protein activity, TFs are often less well correlated to their targets
than the targets are between each other (Figure~\ref{fig:data}-B3).
For
STE12, which regulates itself, all
correlation coefficients exhibit a strong trend towards high
positive correlation. Whereas MCM1, which is not self-regulating,  is far less
strongly correlated to its targets than the targets are between each
other. 
Thus, in general it is more
informative to use the correlation between targets for inference,
which is consistently high whether or not a TF is transcriptionally
regulated itself.  


\para{Gene perturbation data}
Paths in the graph are visible in cause-effect datasets \citep{Hughes2000,Roberts2000}.
We find only very small effects
of perturbations in the pathway on the expression of members of the
membrane and cytoplasm compartment including TFs.
Figure~\ref{fig:data}-C summarizes this result for the
\citet{Roberts2000} data. Very similar results were found for the
\citet{Hughes2000} data. The four boxes correspond to the three
compartments plus TFs. In each box, a vertical line corresponds to a
perturbation in the pathway (some replicated). The dots show the
fold-changes of the pathway genes in this compartment. 
Only in the nuclear
compartment are wide-spread large fold-changes visible. 
This observation motivates the construction of our
likelihood around the presence of paths between the knockout and genes
in the nuclear compartment (see section \ref{sec:model}).  In this way, when the knockout is far
enough upstream, there is information about edges in the cytoplasm as well, even if the proteins there show no effect on the  transcriptional level.



\para{Protein phosphorylation}
A first example of node information is protein phosphorylation.
The study of \citet{Gruhler2005} assessed differential
phosphorylation of proteins in response to pheromone.
Figure~\ref{fig:data}-D1 shows the log-ratios between the pheromone
treated and untreated conditions. Almost all proteins of the
pheromone pathway measured by \citet{Gruhler2005} are up-regulated,
which makes sense for a kinase cascade. The phosphorylation we
observe for proteins corresponding to genes only attributed to the
nuclear compartment in our model must be due to other kinase
pathways in the cell. We further assessed to what extent the
differential phosphorylation is correlated with the pathway model by
fitting an auto-logistic regression. As a
measure of correlation we computed the variance explained,
$R^2=0.76$, using the bootstrap 
. The variance
explained by the auto-logistic regression was found statistically
significant, when compared to the correlation of differential
phosphorylation with randomized pathway models, $p \approx 0.062$,
and with randomized protein permutations on the true pathway model,
$p \approx 0.059$.


\para{Protein domains}
A second example of node information are protein domains.
We retrieved protein domains from PFAM \citep{Punta2012} and SMART
\citep{Letunic2012}. First, we sought to quantify which domains, if
any, were over-represented in the set of proteins involved in the
complete pheromone response pathway as well as in each compartment,
in turn. 
Figure \ref{fig:data}-D2 lists the domains that were found to be
over-represented in the complete pathway and in the cytoplasm;
darker shades of gray indicate a more significant p-value for the
over-representation test. 

Second, we sought to quantify to what extent the presence or absence
of specific protein domains in proteins interacting with a given
protein, $P$, was informative about the presence or absence of the
same domain in such protein, $P$. This analysis was carried out
using auto-logistic models, which
summarize the informativeness of protein domains between interacting
proteins on average, across all proteins in a given pathway. We fit
auto-logistic regressions using each protein $P$ in the
cytoplasm compartment of the pheromone response pathway as data
point, and the presence or absence of domains $D_{1:K}$ in any one
protein among those interacting with $P$ as covariates. 

We fit multivariate models, which assume that the
presence or absence of either the same or complementary domains is a
factor that facilitates protein physical interactions. 
The two tables in \ref{fig:data}-D3 summarize the goodness of fit of
the multivariate models, and report bootstrap p-values to assess the
significance of the AIC scores.  Figure
\ref{fig:data}-D3 shows the p-values obtained by fitting the
multivariate auto-logistic regression to randomized pathway models.
The domains identified by the multivariate
models as putatively carrying signal about the pheromone pathway in
the cytoplasm overlap with the domains identified by the
over-representation analysis above; namely, P21 rho-binding domains,
S-TKc domains, and tyrosine-specific catalytic domains. 

%
%

In summary, node attributes of the proteins involved in the
pheromone response pathways are informative about mechanistic
elements of the kinase cascade, across cellular localizations and in
the cytoplasm. These findings suggest that integrating node
attributes such as protein domains and cellular localization should
increase the likelihood of pathway models that encode real
biological signal about the inner working of a target pathway.

\para{Data Integration} 
The previous results suggest that some
datasets are indeed more informative in certain cellular locations.  For example, protein interactions can
\begin{table}[b!]
\center
{
\caption{The compartment map, $\mathcal{C}$, associates pathway
compartments with those data types that are informative
for such compartments. Prior information is informative for all
compartments. \label{tab:confidence-map}}
\begin{tabular}{rccc}
       & Membrane & Cytoplasm & Nucleus\\ \hline
   PPI & 1        & 1         & 0      \\
    TF & 0        & 0         & 1      \\
  Expr & 0        & 0         & 1      \\
  Kout & 0        & 1         & 1      \\
  Node & 0        & 1         & 0      \\ \hline
 Prior & 1        & 1         & 1      \\ \hline
\end{tabular}}
\label{tab:cmap}
\end{table}
explain wide parts of the kinase cascade in the cytoplasm, while
co-expression is very strong for TF targets. However, no dataset is
informative in all compartments: Neither protein interactions nor
knockout data can explain a complete pathway. The pheromone
response pathway is an archetypical MAPK pathway, so we expect these
observations also to be valid for other MAPK and signaling
pathways. These results suggest that the compartment-specific modeling approach we take
here is sensible.  As a proof of concept, we use the results of
exploratory data analysis to heuristically construct the compartment
map, $\mathcal{C}$ (Table \ref{tab:confidence-map}).
Ultimately, we hope to infer the compartment map in a
statistically principled way.

\subsection{Validation of the integrated analysis}
\label{sec:predict}

We evaluated how well the joint model, which combines all the complementary data
types discussed above, supports the pathway hypothesis in Section \ref{sec:hypothesis} by
sampling 1000 possible pathways using MCMC and tabulating the posterior probabilities over the edges.  

The logistic regression model for domain data may be subject to over-fitting and separation. 
This can occur since there are many different protein domains present, yet the frequency of any single domain is fairly low.  
To mitigate this issue, we used a Cauchy prior on the coefficients for the suto-logistic regression, which is a sensible default prior for this model \citep{Gelman2008}.  Since the domain information in the pheromone pathway is relatively sparse, we also collected protein domain data from other MAPK pathways and used the hypothesized structure of those pathways to help learn the regression coefficients.  Figure \ref{fig:data}A includes the posterior probabilities for the edges in our initial hypothesis.


We also used a {\em leave-one-out} 
\begin{table}[b!]
\caption{ Posterior edge probabilities for leave-one-out trials involving edges in knockout experiments.  
Since we use a leave-node-out scheme, there are two
  posterior probabilities for an edge (corresponding to which of the
  two node endpoints were left out for that particular simulation). 
  \label{tab:postprobs-cv}}
\scriptsize
\begin{center}
\begin{tabular}{|l|ccc|ccc|}
  \hline
&\multicolumn{3}{c|}{Real data}&\multicolumn{3}{c|}{In Silico} \\
\hline 
 & Min & Average & Max & Min & Average & Max\\ 
  \hline
  STE11/STE7 & 0.01 & 0.01 & 0.01 & 0.26 & .31 & 0.36 \\ 
  MCM1/STE2 & 0.00 & 0.01 & 0.02 & 0.03 & 0.12 & 0.2\\ 
  MF(ALPHA)1/STE2 & 0.00 & 0.00 & 0.01 & 0.01 & 0.19 & 0.36\\ 
  FUS1/STE12 & 0.80 & 0.83 & 0.87 & 0.39 & 0.66 & 0.92\\ 
  CDC42/STE18 & 0.00 & 0.00 & 0.00 & 0.00 & 0.16 & 0.31 \\ 
  FUS3/STE12 & 0.01 & 0.01 & 0.01 & 0.01 & 0.10 & 0.19\\ 
  STE5/STE7 & 0.13 & 0.13 & 0.13 & 0.00 & 0.14 & 0.27 \\ 
  BNI1/CDC42 & 0.49 & 0.55 & 0.61 &0.20 & 0.24 &0.28\\ 
  FAR1/MCM1 & 0.00 & 0.00 & 0.00 & 0.24 & 0.26 & 0.27\\ 
  FAR1/STE12 & 0.00 & 0.00 & 0.00 & 0.00 & 0.37 &0.73\\ 
  STE12/CHS1 & 0.80 & 0.82 & 0.83 & 0.01 & 0.02 & 0.03 \\ 
  STE12/FIG2 & 0.84 & 0.84 & 0.85 & 0.04 & 0.24 & 0.43\\ 
  MCM1/AGA1 & 0.10 & 0.23 & 0.37 & 0.07 & 0.17 & 0.27\\ 
  STE12/FIG1 & 0.00 & 0.00 & 0.00 & 0.42 & 0.70 & 0.98\\ 
  STE12/CIK1 & 0.83 & 0.84 & 0.85 & 0.94 & 0.96 & 0.98\\ 
  STE12/KAR5 & 0.83 & 0.83 & 0.84 & 0.23 & 0.30 & 0.37 \\ 
  STE12/GIC2 & 0.83 & 0.83 & 0.84 & 0.12 & 0.54 & 0.95\\ 
  MCM1/SWI4 & 0.00 & 0.00 & 0.00 & 0.16 & 0.29 & 0.41 \\ 
   \hline
\end{tabular}
\end{center}
\end{table}
strategy to evaluate the predictive
power of our model.  We ran 37 separate simulations where each
node was in turn left out of the training pathway. The edges
connected to this node were propagated to the neighboring nodes of the
left-out node.  
We left out the nodes rather than edges, because
specifically leaving out edges is equivalent to assuming that we
know there is no edge present. We needed to construct our model in a way
 that encodes ignorance
about the presence of an edge.  
Leaving out the nodes, instead of
the edges, is one way of being agnostic about the presence of edges attached to that node.  
Only the coefficients in the auto-logistic regression
were learned from the pathway hypothesis, so only the node
likelihoods were affected.  Table \ref{tab:postprobs-cv} shows the
posterior probabilities for edges (under simulations in which a
node was removed from the prior hypothesis pathway).  This table
presents posterior probabilities for edges involved in knockout experiments.

\begin{figure}[ht!]
  \centering
  {\includegraphics[width=0.475\columnwidth]{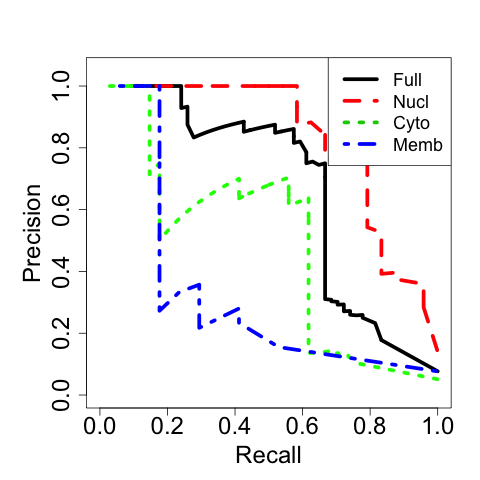}}        
  {\includegraphics[width=0.475\columnwidth]{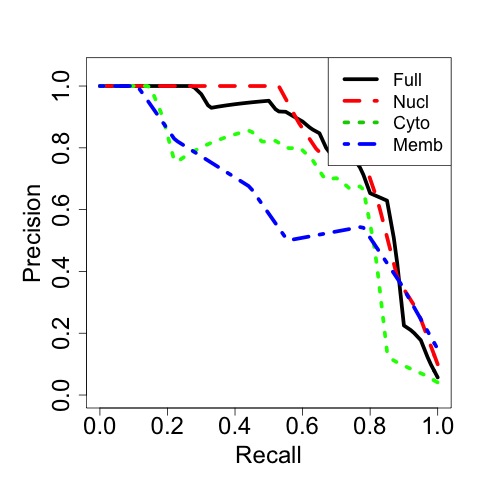}}        
  \caption{\footnotesize Precision/Recall curves overall and by
    compartment for the MAPK pathway (left) and simulated data
    (right).  In truth, the membrane compartment, which has the fewest
    genes, performs poorly because only the PPI dataset is (weakly)
    informative there. The simulated data curve reflects the average
    Precision/Recall over 30 simulated datasets.  \label{pr_full}}
\end{figure}

Lastly, Figure \ref{pr_full} shows the precision-recall curve for our
model, by compartment.  For the membrane compartment, only the PPI
data is informative, and weakly so.  Thus, it performs the most
poorly, although there are also by far the fewest genes in this
compartment.  By contrast, the nuclear and cytoplasm compartments both
have high precision and recall.

\subsection{Inferring cross-talk with other pathways}

With our model, we are also able to identify possible cross-talk
between pathways.  In this paper, we focus on the pheromone response
pathway, but our model can easily be used on other pathways, as long
as we specify the relevant genes and transcription factors, and their
corresponding cellular locations.  

For instance, the MAPK pathway consists of the pheromone sub-pathway,
as well as hypotonic shock, osmolarity and starvation sub-pathways.
The degree of interaction between components of these MAPK pathways is
not currently known.  To identify cross-talk between the pheromone
pathway and other MAPK pathways, we can simply include a new set of
genes from the other sub pathways and fit the model as usual. 
The results for the cross-talk evaluations are displayed in Table \ref{tab:crosstalk}.

\begin{table}[ht!]
\caption{ Number of inferred edges between the pheromone pathway and
  one of the other three sub-pathways with posterior probabilities above
  0.3.  
  \label{tab:crosstalk}}
\begin{center}
\begin{tabular}{lccc}
  \hline
 & osmolarity & hypotonic & starvation \\ 
  \hline
cytoplasm-cytoplasm & 16 & 25 & 11 \\ 
  cytoplasm-membrane & 12 & 17 & 8 \\ 
  cytoplasm-nucleus & 22 & 17 & 3 \\ 
  cytoplasm-tf & 0 & 2 & 3 \\ 
  membrane-membrane & 2 & 2 & 2 \\ 
  membrane-nucleus & 19 & 13 & 3 \\ 
  membrane-tf & 0 & 1 & 2 \\ 
  nucleus-nucleus & 4 & 7 & 0 \\ 
  nucleus-tf & 1 & 6 & 10 \\ 
  tf-tf & 0 & 0 & 2 \\ 
   \hline
\end{tabular}
\end{center}
\end{table}

\subsection{Performance assessment on simulated data}

We also fit the model to \textit{in silico} data.  We constructed the
``true pathway'' to match the hypothesized MAPK pheromone pathway of
Figure \ref{fig:data}A.  That is, we fixed a pathway with the matching
nodes and edges.  We then generated in silico datasets from the models
specified in Section \ref{sec:model}.  The one exception is the data
generation for the node data.  

Here, we generate the presence of domains in a way such that short
chains in the pathway are more likely to share domains than are random
non-neighboring nodes.  Specifically, we randomly chose chains of
length 1 to 4 and added a common ``domain'' to every node in that
chain.  In this way, the domain data realistically reflect the notion
that genes sharing common protein domains are more likely to interact.

The leave-one-out results are given in Table \ref{tab:postprobs-cv}
beside the results for the true data.  Figure \ref{pr_full} shows the
precision-recall curve averaged over 30 simulated datasets.  As in the
true data analysis, the results demonstrate high precision and recall,
expecially in the ``nucleus'' and ``cytoplasm''.  The ``membrane''
shows the worst precision-recall because we have the fewest
informative data types there, but when simulating from the true data
generating process, we still do quite well.

\section{Discussion}  
\begin{figure}[b!]
 \includegraphics[width=0.85\textwidth]{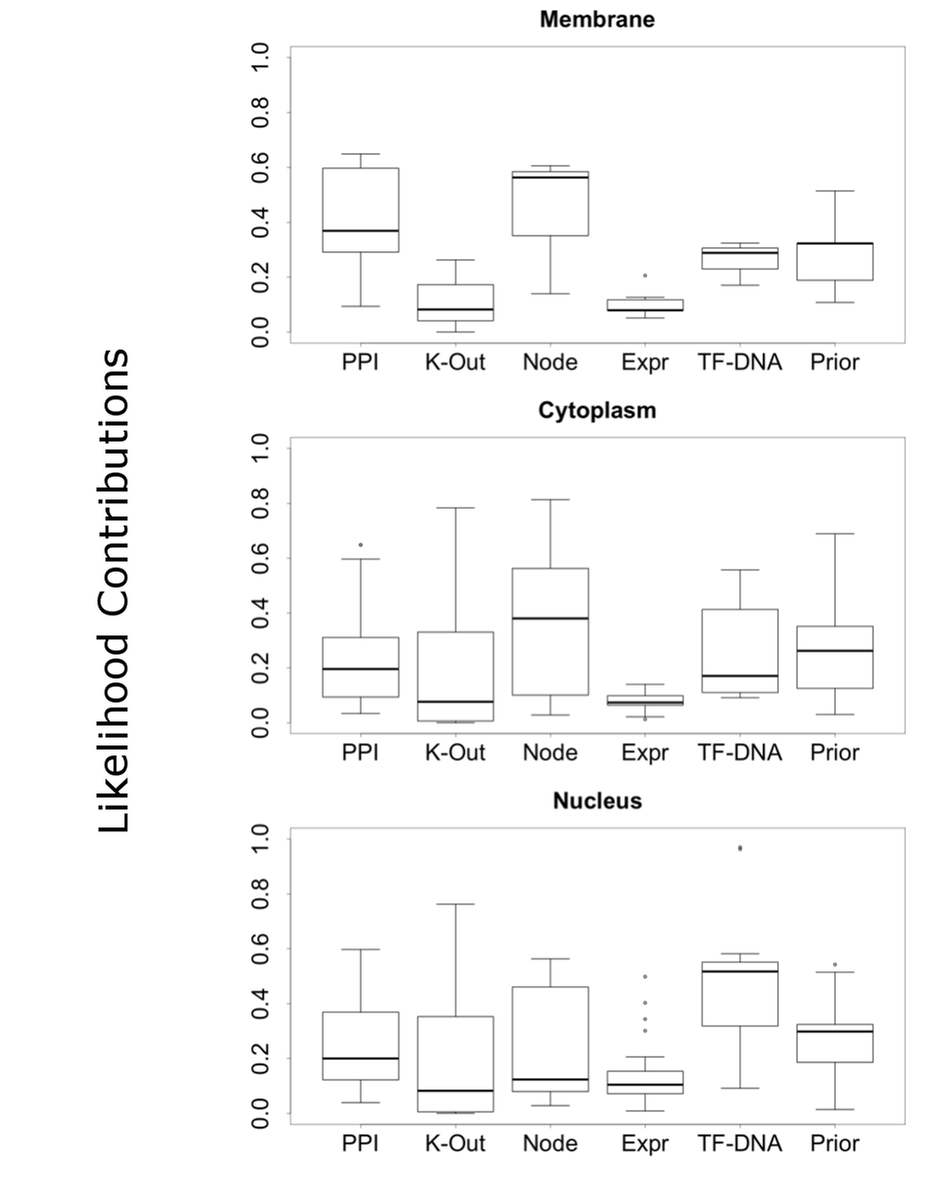}                
  \caption{Percentages of differential likelihood (presence vs. absence of an edge) due to
    specific data types, by compartment. Node
    data contribute the most in the cytoplasm (center), whereas TF-DNA
    binding data contribute the most in the nucleus (right).   \label{fig:contrbs} }
\end{figure}

The proposed methodology achieves fairly strong predictive power by integrating data in a compartment specific way. 
Importantly, we are able to evaluate how each data type
contributes to the overall likelihood of any edge.  Since each data
type independently contributes to the probability of an edge, we can
compute the fraction of the overall likelihood difference (between an edge and
no edge) that is due to a particular data type.  
In this way our framework provides  information about which parts of a pathway hypothesis are not well supported by available data (see Figure \ref{fig:contrbs}).  

%

In addition, our methodology can identify if a particular data type tends to 
disagree with the other data types for sets of edges.  This could indicate whether or
not a data type is at all useful for modeling edges in a particular
cellular location.  Thus, it may be possible to do inference on the compartment map from Table \ref{tab:confidence-map}, rather than fix it
a priori.  Alternatively, this information can be used to check the
validity of the individual data models of Section \ref{sec:model} .  

There are some open statistical issues that could be addressed in future work.  
One problem with the node data, is that the protein domains are
diverse and sparse.  While there is evidence of signal here, there is
an over-fitting problem.  With more domain data, or perhaps broader domain
categories, we may be able to learn more from the prior pathway.  If
this was the case, the leave-one-out results in the cytoplasm might
improve significantly.  This is evident from our results which show how borrowing domain
information from other MAPK sub-pathways significantly improved the
posterior probabilities of edges in the leave-one-out simulations.

We also noticed that most of the knockouts in the gene
perturbation data set we used were generally downstream.  If the
knockouts were further upstream from perturbed genes in the nucleus,
then we could learn about the possible presence of edges in a path
between the knockout and other genes.


Lastly, we divided the pathway into its three main
compartments: membrane, cytoplasm and nucleus.  However, 
in future work, we hope to divide the pathway more finely into the over two dozen cellular components specified by the gene ontology (GO) for the yeast \textit{S. Cerevisae}.  By
dividing the pathway into more compartments, we would also have a
greater degree of control over which data types are used in various
parts of the cell.  

\subsection{Concluding remarks} In this paper we introduced a technique for
refining cellular pathway models by integrating heterogeneous data sources in a compartment specific way and explicitly included node properties in our model.   
Our case-study results indicate that this model can be useful for
discovering new components or cross-talk with other pathways.
Our powerful and flexible pathway modeling framework can be easily
extended and modified to include additional and novel datasets.

\appendix
\section{Supplementary results}

In this appendix we present more details about the simulation results.

\begin{figure}[ht!]
  \centering
  {\includegraphics[width=0.7\columnwidth]{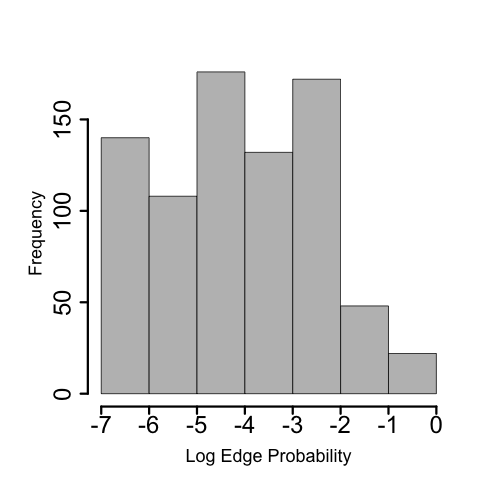}}        
  \caption{\footnotesize Log posterior probabilities for edges that were
    not in the hypothesis pathway.  The vast majority of non-edges
    have small posterior probability (third quantile at 0.02).
    However, there are a few highly probable edges, which may
    indicate previously undiscovered interactions.  \label{noEdge}}
\end{figure}
\clearpage

\begin{table}[ht!]
\begin{minipage}{0.45\textwidth}
\scriptsize
\raggedright
\begin{tabular}{rllr}
  \hline
 & Gene 1 & Gene 2 & Prob \\
  \hline
1 & STE12 & DIG2 & 0.60 \\
  2 & STE12 & FUS1 & 0.39 \\
  3 & STE12 & FUS3 & 0.01 \\
  4 & STE12 & FAR1 & 0.00 \\
  5 & STE12 & MCM1 & 0.00 \\
  6 & STE12 & FIG2 & 0.43 \\
  7 & STE12 & FIG1 & 0.42 \\
  8 & STE12 & CIK1 & 0.98 \\
  9 & STE12 & GIC2 & 0.12 \\
  10 & STE12 & AFR1 & 0.01 \\
  11 & STE12 & KAR5 & 0.23 \\
  12 & STE12 & CHS1 & 0.03 \\
  13 & STE12 & AGA1 & 0.27 \\
  14 & DIG2 & STE12 & 0.68 \\
  15 & DIG2 & FUS3 & 0.00 \\
  16 & STE7 & STE11 & 0.26 \\
  17 & STE7 & STE5 & 0.21 \\
  18 & STE7 & FUS3 & 0.26 \\
  19 & STE11 & STE7 & 0.36 \\
  20 & STE11 & STE20 & 0.24 \\
  21 & STE11 & STE5 & 0.00 \\
  22 & STE20 & STE11 & 0.00 \\
  23 & STE20 & CDC42 & 0.31 \\
  24 & STE20 & BEM1 & 0.08 \\
  25 & STE20 & STE5 & 0.00 \\
  26 & CDC42 & STE20 & 0.00 \\
  27 & CDC42 & BNI1 & 0.28 \\
  28 & CDC42 & STE4 & 0.24 \\
  29 & CDC42 & STE18 & 0.31 \\
  30 & CDC42 & BEM1 & 0.47 \\
  31 & CDC42 & CDC24 & 0.50 \\
  32 & FUS1 & STE12 & 0.98 \\
  33 & BNI1 & CDC42 & 0.20 \\
  34 & MFA1 & STE3 & 0.34 \\
  35 & MFA1 & MCM1 & 0.07 \\
  36 & STE2 & MF(ALPHA)2 & 0.01 \\
  37 & STE2 & GPA1 & 0.35 \\
  38 & STE2 & MCM1 & 0.20 \\
  39 & STE3 & MFA1 & 0.30 \\
  40 & STE3 & GPA1 & 0.13 \\
  41 & MF(ALPHA)2 & STE2 & 0.36 \\
  42 & GPA1 & STE2 & 0.01 \\
  43 & GPA1 & STE3 & 0.14 \\
  44 & GPA1 & STE4 & 0.14 \\
  45 & GPA1 & STE18 & 0.12 \\
  46 & STE4 & CDC42 & 0.22 \\
  47 & STE4 & GPA1 & 0.14 \\
  48 & STE18 & CDC42 & 0.00 \\
  49 & STE18 & GPA1 & 0.13 \\
  50 & BEM1 & STE20 & 0.36 \\
  51 & BEM1 & CDC42 & 0.18 \\
  52 & CDC24 & CDC42 & 0.18 \\
  53 & STE5 & STE7 & 0.00 \\
  54 & STE5 & STE11 & 0.00 \\
  55 & STE5 & STE20 & 0.00 \\
  56 & STE5 & FUS3 & 0.00 \\
  57 & FUS3 & STE12 & 0.19 \\
  58 & FUS3 & DIG2 & 0.21 \\
  59 & FUS3 & STE7 & 0.22 \\
  60 & FUS3 & STE5 & 0.05 \\
   \hline
\end{tabular}
\end{minipage}
\begin{minipage}{.45\textwidth}
\scriptsize
\raggedleft
\vspace{-4in}
\begin{tabular}{rllr}
  \hline
 & Gene 1 & Gene 2 & Prob \\
  \hline
  61 & FUS3 & MSG5 & 0.05 \\
  62 & FUS3 & FAR1 & 0.00 \\
  63 & MSG5 & FUS3 & 0.00 \\
  64 & FAR1 & STE12 & 0.73 \\
  65 & FAR1 & FUS3 & 0.27 \\
  66 & FAR1 & MCM1 & 0.27 \\
  67 & MCM1 & STE12 & 0.00 \\
  68 & MCM1 & MFA1 & 0.15 \\
  69 & MCM1 & STE2 & 0.03 \\
  70 & MCM1 & FAR1 & 0.24 \\
  71 & MCM1 & SWI4 & 0.41 \\
  72 & MCM1 & MFA2 & 0.20 \\
  73 & MCM1 & AGA1 & 0.27 \\
  74 & MCM1 & ALK1 & 0.15 \\
  75 & MCM1 & SWI5 & 0.38 \\
  76 & MCM1 & CDC20 & 0.34 \\
  77 & SWI4 & MCM1 & 0.16 \\
  78 & MFA2 & MCM1 & 0.19 \\
  79 & FIG2 & STE12 & 0.04 \\
  80 & FIG1 & STE12 & 0.98 \\
  81 & CIK1 & STE12 & 0.94 \\
  82 & GIC2 & STE12 & 0.95 \\
  83 & AFR1 & STE12 & 0.02 \\
  84 & KAR5 & STE12 & 0.37 \\
  85 & CHS1 & STE12 & 0.01 \\
  86 & AGA1 & STE12 & 0.00 \\
  87 & AGA1 & MCM1 & 0.07 \\
  88 & ALK1 & MCM1 & 0.24 \\
  89 & SWI5 & MCM1 & 0.13 \\
  90 & CDC20 & MCM1 & 0.18 \\
   \hline
\end{tabular}
\end{minipage}
\caption{Posterior edge probabilities. \label{tab:edge-probs}}
\end{table}
\clearpage

\bibliographystyle{chicago}
\bibliography{allrefs2}

\begin{thebibliography}{}

\bibitem[\protect\citeauthoryear{Balbin, Prensner, Sahu, Yocum, Shankar, Malik,
  Fermin, Dhanasekaran, Chandler, Thomas, Beer, Cao, Nesvizhskii, and
  Chinnaiyan}{Balbin et~al.}{2013}]{Balbin2013}
Balbin, O.~A., J.~R. Prensner, A.~Sahu, A.~Yocum, S.~Shankar, R.~Malik,
  D.~Fermin, S.~M. Dhanasekaran, B.~Chandler, D.~Thomas, D.~G. Beer, X.~Cao,
  A.~I. Nesvizhskii, and A.~M. Chinnaiyan (2013, Oct).
\newblock Reconstructing targetable pathways in lung cancer by integrating
  diverse omics data.
\newblock {\em Nat Commun\/}~{\em 4}, 2617.

\bibitem[\protect\citeauthoryear{Bernard and Hartemink}{Bernard and
  Hartemink}{2005}]{Bernard2005}
Bernard, A. and A.~J. Hartemink (2005).
\newblock Informative structure priors: joint learning of dynamic regulatory
  networks from multiple types of data.
\newblock {\em Pac Symp Biocomput\/}, 459--470.

\bibitem[\protect\citeauthoryear{Brem and Kruglyak}{Brem and
  Kruglyak}{2005}]{Brem2005}
Brem, R.~B. and L.~Kruglyak (2005, Feb).
\newblock The landscape of genetic complexity across 5,700 gene expression
  traits in yeast.
\newblock {\em Proc Natl Acad Sci U S A\/}~{\em 102\/}(5), 1572--1577.

\bibitem[\protect\citeauthoryear{Friedman}{Friedman}{2004}]{Friedman2004}
Friedman, N. (2004, Feb).
\newblock Inferring cellular networks using probabilistic graphical models.
\newblock {\em Science\/}~{\em 303\/}(5659), 799--805.

\bibitem[\protect\citeauthoryear{Fr{\"o}hlich, Beissbarth, Tresch, Kostka,
  Jacob, Spang, and Markowetz}{Fr{\"o}hlich et~al.}{2008}]{Froehlich2008a}
Fr{\"o}hlich, H., T.~Beissbarth, A.~Tresch, D.~Kostka, J.~Jacob, R.~Spang, and
  F.~Markowetz (2008, Nov).
\newblock Analyzing gene perturbation screens with nested effects models in r
  and bioconductor.
\newblock {\em Bioinformatics\/}~{\em 24\/}(21), 2549--2550.

\bibitem[\protect\citeauthoryear{Fr{\"o}hlich, Fellmann, S{\"u}ltmann, Poustka,
  and Bei{\ss}barth}{Fr{\"o}hlich et~al.}{2007}]{Froehlich2007}
Fr{\"o}hlich, H., M.~Fellmann, H.~S{\"u}ltmann, A.~Poustka, and
  T.~Bei{\ss}barth (2007).
\newblock Large scale statistical inference of signaling pathways from rnai and
  microarray data.
\newblock {\em BMC Bioinformatics\/}~{\em 8}, 386.

\bibitem[\protect\citeauthoryear{Fr{\"o}hlich, Fellmann, S{\"u}ltmann, Poustka,
  and Bei{\ss}barth}{Fr{\"o}hlich et~al.}{2008}]{Froehlich2008}
Fr{\"o}hlich, H., M.~Fellmann, H.~S{\"u}ltmann, A.~Poustka, and
  T.~Bei{\ss}barth (2008, Oct).
\newblock Predicting pathway membership via domain signatures.
\newblock {\em Bioinformatics\/}~{\em 24\/}(19), 2137--2142.

\bibitem[\protect\citeauthoryear{Gasch, Spellman, Kao, Carmel-Harel, Eisen,
  Storz, Botstein, and Brown}{Gasch et~al.}{2000}]{Gasch2000}
Gasch, A.~P., P.~T. Spellman, C.~M. Kao, O.~Carmel-Harel, M.~B. Eisen,
  G.~Storz, D.~Botstein, and P.~O. Brown (2000, Dec).
\newblock Genomic expression programs in the response of yeast cells to
  environmental changes.
\newblock {\em Mol Biol Cell\/}~{\em 11\/}(12), 4241--4257.

\bibitem[\protect\citeauthoryear{Gat-Viks and Shamir}{Gat-Viks and
  Shamir}{2007}]{Gat-Viks2007}
Gat-Viks, I. and R.~Shamir (2007, Mar).
\newblock Refinement and expansion of signaling pathways: the osmotic response
  network in yeast.
\newblock {\em Genome Res\/}~{\em 17\/}(3), 358--367.

\bibitem[\protect\citeauthoryear{Gelman}{Gelman}{2008}]{Gelman2008}
Gelman, A. (2008).
\newblock A weakly informative default prior distribution for logistic and
  other regression models.
\newblock {\em The Annals of Applied Statistics\/}~{\em 2\/}(4), 1360--1383.

\bibitem[\protect\citeauthoryear{Gitter, Carmi, Barkai, and Bar-Joseph}{Gitter
  et~al.}{2013}]{Gitter2013}
Gitter, A., M.~Carmi, N.~Barkai, and Z.~Bar-Joseph (2013, Feb).
\newblock Linking the signaling cascades and dynamic regulatory networks
  controlling stress responses.
\newblock {\em Genome Res\/}~{\em 23\/}(2), 365--376.

\bibitem[\protect\citeauthoryear{Gruhler, Olsen, Mohammed, Mortensen,
  Faergeman, Mann, and Jensen}{Gruhler et~al.}{2005}]{Gruhler2005}
Gruhler, A., J.~V. Olsen, S.~Mohammed, P.~Mortensen, N.~J. Faergeman, M.~Mann,
  and O.~N. Jensen (2005, Mar).
\newblock Quantitative phosphoproteomics applied to the yeast pheromone
  signaling pathway.
\newblock {\em Mol Cell Proteomics\/}~{\em 4\/}(3), 310--327.

\bibitem[\protect\citeauthoryear{Guan, Gorenshteyn, Burmeister, Wong,
  Schimenti, Handel, Bult, Hibbs, and Troyanskaya}{Guan
  et~al.}{2012}]{Guan2012}
Guan, Y., D.~Gorenshteyn, M.~Burmeister, A.~K. Wong, J.~C. Schimenti, M.~A.
  Handel, C.~J. Bult, M.~A. Hibbs, and O.~G. Troyanskaya (2012).
\newblock Tissue-specific functional networks for prioritizing phenotype and
  disease genes.
\newblock {\em PLoS Comput Biol\/}~{\em 8\/}(9), e1002694.

\bibitem[\protect\citeauthoryear{Guan, Myers, Hess, Barutcuoglu, Caudy, and
  Troyanskaya}{Guan et~al.}{2008}]{Guan2008a}
Guan, Y., C.~L. Myers, D.~C. Hess, Z.~Barutcuoglu, A.~A. Caudy, and O.~G.
  Troyanskaya (2008).
\newblock Predicting gene function in a hierarchical context with an ensemble
  of classifiers.
\newblock {\em Genome Biol\/}~{\em 9 Suppl 1}, S3.

\bibitem[\protect\citeauthoryear{Hahne, Mehrle, Arlt, Poustka, Wiemann, and
  Bei{\ss}barth}{Hahne et~al.}{2008}]{Hahne2008}
Hahne, F., A.~Mehrle, D.~Arlt, A.~Poustka, S.~Wiemann, and T.~Bei{\ss}barth
  (2008).
\newblock Extending pathways based on gene lists using {InterPro} domain
  signatures.
\newblock {\em BMC Bioinformatics\/}~{\em 9}, 3.

\bibitem[\protect\citeauthoryear{Hara, Ono, Kuroda, and Ueda}{Hara
  et~al.}{2012}]{Hara2012}
Hara, K., T.~Ono, K.~Kuroda, and M.~Ueda (2012, May).
\newblock Membrane-displayed peptide ligand activates the pheromone response
  pathway in saccharomyces cerevisiae.
\newblock {\em J Biochem\/}~{\em 151\/}(5), 551--557.

\bibitem[\protect\citeauthoryear{Harbison, Gordon, Lee, Rinaldi, Macisaac,
  Danford, Hannett, Tagne, Reynolds, Yoo, Jennings, Zeitlinger, Pokholok,
  Kellis, Rolfe, Takusagawa, Lander, Gifford, Fraenkel, and Young}{Harbison
  et~al.}{2004}]{Harbison2004}
Harbison, C.~T., D.~B. Gordon, T.~I. Lee, N.~J. Rinaldi, K.~D. Macisaac, T.~W.
  Danford, N.~M. Hannett, J.-B. Tagne, D.~B. Reynolds, J.~Yoo, E.~G. Jennings,
  J.~Zeitlinger, D.~K. Pokholok, M.~Kellis, P.~A. Rolfe, K.~T. Takusagawa,
  E.~S. Lander, D.~K. Gifford, E.~Fraenkel, and R.~A. Young (2004, Sep).
\newblock Transcriptional regulatory code of a eukaryotic genome.
\newblock {\em Nature\/}~{\em 431\/}(7004), 99--104.

\bibitem[\protect\citeauthoryear{Hibbs, Hess, Myers, Huttenhower, Li, and
  Troyanskaya}{Hibbs et~al.}{2007}]{Hibbs2007}
Hibbs, M.~A., D.~C. Hess, C.~L. Myers, C.~Huttenhower, K.~Li, and O.~G.
  Troyanskaya (2007, Oct).
\newblock Exploring the functional landscape of gene expression: directed
  search of large microarray compendia.
\newblock {\em Bioinformatics\/}~{\em 23\/}(20), 2692--2699.

\bibitem[\protect\citeauthoryear{Hibbs, Myers, Huttenhower, Hess, Li, Caudy,
  and Troyanskaya}{Hibbs et~al.}{2008}]{Hibbs2008}
Hibbs, M.~A., C.~L. Myers, C.~Huttenhower, D.~C. Hess, K.~Li, A.~A. Caudy, and
  O.~G. Troyanskaya (2008).
\newblock Analysis of computational functional genomic approaches for directing
  experimental biology: a case study in mitochondrial inheritance.
\newblock {\em PLoS Comput Biol\/}~{\em in press}.

\bibitem[\protect\citeauthoryear{Hughes, Marton, Jones, Roberts, Stoughton,
  Armour, Bennett, Coffey, Dai, He, Kidd, King, Meyer, Slade, Lum, Stepaniants,
  Shoemaker, Gachotte, Chakraburtty, Simon, Bard, and Friend}{Hughes
  et~al.}{2000}]{Hughes2000}
Hughes, T.~R., M.~J. Marton, A.~R. Jones, C.~J. Roberts, R.~Stoughton, C.~D.
  Armour, H.~A. Bennett, E.~Coffey, H.~Dai, Y.~D. He, M.~J. Kidd, A.~M. King,
  M.~R. Meyer, D.~Slade, P.~Y. Lum, S.~B. Stepaniants, D.~D. Shoemaker,
  D.~Gachotte, K.~Chakraburtty, J.~Simon, M.~Bard, and S.~H. Friend (2000,
  Jul).
\newblock Functional discovery via a compendium of expression profiles.
\newblock {\em Cell\/}~{\em 102\/}(1), 109--126.

\bibitem[\protect\citeauthoryear{Hyduke and Palsson}{Hyduke and
  Palsson}{2010}]{Hyduke2010}
Hyduke, D.~R. and B.~Ø. Palsson (2010, Apr).
\newblock Towards genome-scale signalling network reconstructions.
\newblock {\em Nat Rev Genet\/}~{\em 11\/}(4), 297--307.

\bibitem[\protect\citeauthoryear{Isci, Dogan, Ozturk, and Otu}{Isci
  et~al.}{2013}]{Isci2013}
Isci, S., H.~Dogan, C.~Ozturk, and H.~H. Otu (2013, Nov).
\newblock Bayesian network prior: network analysis of biological data using
  external knowledge.
\newblock {\em Bioinformatics\/}.

\bibitem[\protect\citeauthoryear{Kanehisa and Goto}{Kanehisa and
  Goto}{2000}]{Kanehisa2000}
Kanehisa, M. and S.~Goto (2000, Jan).
\newblock Kegg: kyoto encyclopedia of genes and genomes.
\newblock {\em Nucleic Acids Res\/}~{\em 28\/}(1), 27--30.

\bibitem[\protect\citeauthoryear{Kirouac, Saez-Rodriguez, Swantek, Burke,
  Lauffenburger, and Sorger}{Kirouac et~al.}{2012}]{Kirouac2012}
Kirouac, D.~C., J.~Saez-Rodriguez, J.~Swantek, J.~M. Burke, D.~A.
  Lauffenburger, and P.~K. Sorger (2012).
\newblock Creating and analyzing pathway and protein interaction compendia for
  modelling signal transduction networks.
\newblock {\em BMC Syst Biol\/}~{\em 6}, 29.

\bibitem[\protect\citeauthoryear{Knapp and Kaderali}{Knapp and
  Kaderali}{2013}]{Knapp2013}
Knapp, B. and L.~Kaderali (2013).
\newblock Reconstruction of cellular signal transduction networks using
  perturbation assays and linear programming.
\newblock {\em PLoS One\/}~{\em 8\/}(7), e69220.

\bibitem[\protect\citeauthoryear{Kofahl and Klipp}{Kofahl and
  Klipp}{2004}]{Kofahl2004}
Kofahl, B. and E.~Klipp (2004, Jul).
\newblock Modelling the dynamics of the yeast pheromone pathway.
\newblock {\em Yeast\/}~{\em 21\/}(10), 831--850.

\bibitem[\protect\citeauthoryear{Letunic, Doerks, and Bork}{Letunic
  et~al.}{2012}]{Letunic2012}
Letunic, I., T.~Doerks, and P.~Bork (2012, Jan).
\newblock Smart 7: recent updates to the protein domain annotation resource.
\newblock {\em Nucleic Acids Res\/}~{\em 40\/}(Database issue), D302--D305.

\bibitem[\protect\citeauthoryear{Li, Wei, Liu, and Zhao}{Li
  et~al.}{2013}]{Li2013}
Li, J., H.~Wei, T.~Liu, and P.~X. Zhao (2013, Oct).
\newblock Gplexus: enabling genome-scale gene association network
  reconstruction and analysis for very large-scale expression data.
\newblock {\em Nucleic Acids Res\/}.

\bibitem[\protect\citeauthoryear{Llewellyn and Eisenberg}{Llewellyn and
  Eisenberg}{2008}]{Llewellyn2008}
Llewellyn, R. and D.~S. Eisenberg (2008, Nov).
\newblock Annotating proteins with generalized functional linkages.
\newblock {\em Proc Natl Acad Sci U S A\/}.

\bibitem[\protect\citeauthoryear{Lo, Raftery, Dombek, Zhu, Schadt, Bumgarner,
  and Yeung}{Lo et~al.}{2012}]{Lo2012}
Lo, K., A.~E. Raftery, K.~M. Dombek, J.~Zhu, E.~E. Schadt, R.~E. Bumgarner, and
  K.~Y. Yeung (2012).
\newblock Integrating external biological knowledge in the construction of
  regulatory networks from time-series expression data.
\newblock {\em BMC Syst Biol\/}~{\em 6}, 101.

\bibitem[\protect\citeauthoryear{Markowetz, Kostka, Troyanskaya, and
  Spang}{Markowetz et~al.}{2007}]{Markowetz2007a}
Markowetz, F., D.~Kostka, O.~G. Troyanskaya, and R.~Spang (2007, Jul).
\newblock Nested effects models for high-dimensional phenotyping screens.
\newblock {\em Bioinformatics\/}~{\em 23\/}(13), i305--i312.

\bibitem[\protect\citeauthoryear{Markowetz and Spang}{Markowetz and
  Spang}{2007}]{Markowetz2007}
Markowetz, F. and R.~Spang (2007).
\newblock Inferring cellular networks--a review.
\newblock {\em BMC Bioinformatics\/}~{\em 8 Suppl 6}, S5.

\bibitem[\protect\citeauthoryear{Mazza, Gat-Viks, Farhan, and Sharan}{Mazza
  et~al.}{2013}]{Mazza2013}
Mazza, A., I.~Gat-Viks, H.~Farhan, and R.~Sharan (2013, July).
\newblock A minimum-labeling approach for reconstructing protein networks
  across multiple conditions.

\bibitem[\protect\citeauthoryear{McClean, Mody, Broach, and Ramanathan}{McClean
  et~al.}{2007}]{McClean2007}
McClean, M.~N., A.~Mody, J.~R. Broach, and S.~Ramanathan (2007, Mar).
\newblock Cross-talk and decision making in {MAP} kinase pathways.
\newblock {\em Nat Genet\/}~{\em 39\/}(3), 409--414.

\bibitem[\protect\citeauthoryear{Mukherjee and Speed}{Mukherjee and
  Speed}{2008}]{Mukherjee2008a}
Mukherjee, S. and T.~P. Speed (2008, Sep).
\newblock {{N}etwork inference using informative priors}.
\newblock {\em Proc. Natl. Acad. Sci. U.S.A.\/}~{\em 105\/}(38), 14313--14318.

\bibitem[\protect\citeauthoryear{Mulder, Wang, Escriu, Ito, Schwarz, Gillis,
  Sirokmány, Donati, Uribe-Lewis, Pavlidis, Murrell, Markowetz, and
  Watt}{Mulder et~al.}{2012}]{Mulder2012}
Mulder, K.~W., X.~Wang, C.~Escriu, Y.~Ito, R.~F. Schwarz, J.~Gillis,
  G.~Sirokmány, G.~Donati, S.~Uribe-Lewis, P.~Pavlidis, A.~Murrell,
  F.~Markowetz, and F.~M. Watt (2012, Jul).
\newblock Diverse epigenetic strategies interact to control epidermal
  differentiation.
\newblock {\em Nat Cell Biol\/}~{\em 14\/}(7), 753--763.

\bibitem[\protect\citeauthoryear{M{\"u}ller, Kuttenkeuler, Gesellchen, Zeidler,
  and Boutros}{M{\"u}ller et~al.}{2005}]{Mueller2005}
M{\"u}ller, P., D.~Kuttenkeuler, V.~Gesellchen, M.~P. Zeidler, and M.~Boutros
  (2005, Aug).
\newblock Identification of {JAK/STAT} signalling components by genome-wide rna
  interference.
\newblock {\em Nature\/}~{\em 436\/}(7052), 871--875.

\bibitem[\protect\citeauthoryear{Myers, Robson, Wible, Hibbs, Chiriac,
  Theesfeld, Dolinski, and Troyanskaya}{Myers et~al.}{2005}]{Myers2005}
Myers, C.~L., D.~Robson, A.~Wible, M.~A. Hibbs, C.~Chiriac, C.~L. Theesfeld,
  K.~Dolinski, and O.~G. Troyanskaya (2005).
\newblock Discovery of biological networks from diverse functional genomic
  data.
\newblock {\em Genome Biol\/}~{\em 6\/}(13), R114.

\bibitem[\protect\citeauthoryear{Nagiec and Dohlman}{Nagiec and
  Dohlman}{2012}]{Nagiec2012}
Nagiec, M.~J. and H.~G. Dohlman (2012, Jan).
\newblock Checkpoints in a yeast differentiation pathway coordinate signaling
  during hyperosmotic stress.
\newblock {\em PLoS Genet\/}~{\em 8\/}(1), e1002437.

\bibitem[\protect\citeauthoryear{Nariai, Kim, Imoto, and Miyano}{Nariai
  et~al.}{2004}]{Nariai2004}
Nariai, N., S.~Kim, S.~Imoto, and S.~Miyano (2004).
\newblock Using protein-protein interactions for refining gene networks
  estimated from microarray data by bayesian networks.
\newblock {\em Pac Symp Biocomput\/}, 336--347.

\bibitem[\protect\citeauthoryear{Ourfali, Shlomi, Ideker, Ruppin, and
  Sharan}{Ourfali et~al.}{2007}]{Ourfali2007}
Ourfali, O., T.~Shlomi, T.~Ideker, E.~Ruppin, and R.~Sharan (2007, Jul).
\newblock {SPINE}: a framework for signaling-regulatory pathway inference from
  cause-effect experiments.
\newblock {\em Bioinformatics\/}~{\em 23\/}(13), i359--i366.

\bibitem[\protect\citeauthoryear{Pounds and Morris}{Pounds and
  Morris}{2003}]{Poun:Morr:2003}
Pounds, S. and S.~W. Morris (2003).
\newblock Estimating the occurrence of false positives and false negatives in
  microarray studies by approximating and partitioning the empirical
  distribution of p-values.
\newblock {\em Bioinformatics\/}~{\em 19\/}(10), 1236--1242.

\bibitem[\protect\citeauthoryear{Punta, Coggill, Eberhardt, Mistry, Tate,
  Boursnell, Pang, Forslund, Ceric, Clements, Heger, Holm, Sonnhammer, Eddy,
  Bateman, and Finn}{Punta et~al.}{2012}]{Punta2012}
Punta, M., P.~C. Coggill, R.~Y. Eberhardt, J.~Mistry, J.~Tate, C.~Boursnell,
  N.~Pang, K.~Forslund, G.~Ceric, J.~Clements, A.~Heger, L.~Holm, E.~L.~L.
  Sonnhammer, S.~R. Eddy, A.~Bateman, and R.~D. Finn (2012, Jan).
\newblock The pfam protein families database.
\newblock {\em Nucleic Acids Res\/}~{\em 40\/}(Database issue), D290--D301.

\bibitem[\protect\citeauthoryear{Reguly, Breitkreutz, Boucher, Breitkreutz,
  Hon, Myers, Parsons, Friesen, Oughtred, Tong, Stark, Ho, Botstein, Andrews,
  Boone, Troyanskya, Ideker, Dolinski, Batada, and Tyers}{Reguly
  et~al.}{2006}]{Reguly2006}
Reguly, T., A.~Breitkreutz, L.~Boucher, B.-J. Breitkreutz, G.~C. Hon, C.~L.
  Myers, A.~Parsons, H.~Friesen, R.~Oughtred, A.~Tong, C.~Stark, Y.~Ho,
  D.~Botstein, B.~Andrews, C.~Boone, O.~G. Troyanskya, T.~Ideker, K.~Dolinski,
  N.~N. Batada, and M.~Tyers (2006).
\newblock Comprehensive curation and analysis of global interaction networks in
  saccharomyces cerevisiae.
\newblock {\em J Biol\/}~{\em 5\/}(4), 11.

\bibitem[\protect\citeauthoryear{Ren, Robert, Wyrick, Aparicio, Jennings,
  Simon, Zeitlinger, Schreiber, Hannett, Kanin, Volkert, Wilson, Bell, and
  Young}{Ren et~al.}{2000}]{Ren2000}
Ren, B., F.~Robert, J.~J. Wyrick, O.~Aparicio, E.~G. Jennings, I.~Simon,
  J.~Zeitlinger, J.~Schreiber, N.~Hannett, E.~Kanin, T.~L. Volkert, C.~J.
  Wilson, S.~P. Bell, and R.~A. Young (2000, Dec).
\newblock Genome-wide location and function of dna binding proteins.
\newblock {\em Science\/}~{\em 290\/}(5500), 2306--2309.

\bibitem[\protect\citeauthoryear{Roberts, Nelson, Marton, Stoughton, Meyer,
  Bennett, He, Dai, Walker, Hughes, Tyers, Boone, and Friend}{Roberts
  et~al.}{2000}]{Roberts2000}
Roberts, C.~J., B.~Nelson, M.~J. Marton, R.~Stoughton, M.~R. Meyer, H.~A.
  Bennett, Y.~D. He, H.~Dai, W.~L. Walker, T.~R. Hughes, M.~Tyers, C.~Boone,
  and S.~H. Friend (2000, Feb).
\newblock Signaling and circuitry of multiple {MAPK} pathways revealed by a
  matrix of global gene expression profiles.
\newblock {\em Science\/}~{\em 287\/}(5454), 873--880.

\bibitem[\protect\citeauthoryear{Ryan, Cimerman?i?, Szpiech, Sali, Hernandez,
  and Krogan}{Ryan et~al.}{2013}]{Ryan2013}
Ryan, C.~J., P.~Cimerman?i?, Z.~A. Szpiech, A.~Sali, R.~D. Hernandez, and N.~J.
  Krogan (2013, Dec).
\newblock High-resolution network biology: connecting sequence with function.
\newblock {\em Nat Rev Genet\/}~{\em 14\/}(12), 865--879.

\bibitem[\protect\citeauthoryear{Sch{\"a}fer and Strimmer}{Sch{\"a}fer and
  Strimmer}{2005a}]{Schaefer2005a}
Sch{\"a}fer, J. and K.~Strimmer (2005a, Mar).
\newblock An empirical {Bayes} approach to inferring large-scale gene
  association networks.
\newblock {\em Bioinformatics\/}~{\em 21\/}(6), 754--764.

\bibitem[\protect\citeauthoryear{Sch{\"a}fer and Strimmer}{Sch{\"a}fer and
  Strimmer}{2005b}]{Schaefer2005}
Sch{\"a}fer, J. and K.~Strimmer (2005b).
\newblock A shrinkage approach to large-scale covariance matrix estimation and
  implications for functional genomics.
\newblock {\em Stat Appl Genet Mol Biol\/}~{\em 4}, Article32.

\bibitem[\protect\citeauthoryear{Schultz, Milpetz, Bork, and Ponting}{Schultz
  et~al.}{1998}]{Schultz1998}
Schultz, J., F.~Milpetz, P.~Bork, and C.~P. Ponting (1998, May).
\newblock Smart, a simple modular architecture research tool: identification of
  signaling domains.
\newblock {\em Proc Natl Acad Sci U S A\/}~{\em 95\/}(11), 5857--5864.

\bibitem[\protect\citeauthoryear{Scott, Ideker, Karp, and Sharan}{Scott
  et~al.}{2006}]{Scott2006}
Scott, J., T.~Ideker, R.~M. Karp, and R.~Sharan (2006, Mar).
\newblock Efficient algorithms for detecting signaling pathways in protein
  interaction networks.
\newblock {\em J Comput Biol\/}~{\em 13\/}(2), 133--144.

\bibitem[\protect\citeauthoryear{Segal, Shapira, Regev, Pe'er, Botstein,
  Koller, and Friedman}{Segal et~al.}{2003}]{Segal2003}
Segal, E., M.~Shapira, A.~Regev, D.~Pe'er, D.~Botstein, D.~Koller, and
  N.~Friedman (2003, Jun).
\newblock Module networks: identifying regulatory modules and their
  condition-specific regulators from gene expression data.
\newblock {\em Nat Genet\/}~{\em 34\/}(2), 166--176.

\bibitem[\protect\citeauthoryear{Segal, Wang, and Koller}{Segal
  et~al.}{2003}]{Segal2003a}
Segal, E., H.~Wang, and D.~Koller (2003).
\newblock Discovering molecular pathways from protein interaction and gene
  expression data.
\newblock {\em Bioinformatics\/}~{\em 19 Suppl 1}, i264--i271.

\bibitem[\protect\citeauthoryear{Simon, Barnett, Hannett, Harbison, Rinaldi,
  Volkert, Wyrick, Zeitlinger, Gifford, Jaakkola, and Young}{Simon
  et~al.}{2001}]{Simon2001}
Simon, I., J.~Barnett, N.~Hannett, C.~T. Harbison, N.~J. Rinaldi, T.~L.
  Volkert, J.~J. Wyrick, J.~Zeitlinger, D.~K. Gifford, T.~S. Jaakkola, and
  R.~A. Young (2001, Sep).
\newblock Serial regulation of transcriptional regulators in the yeast cell
  cycle.
\newblock {\em Cell\/}~{\em 106\/}(6), 697--708.

\bibitem[\protect\citeauthoryear{Stark, Breitkreutz, Reguly, Boucher,
  Breitkreutz, and Tyers}{Stark et~al.}{2006}]{Stark2006}
Stark, C., B.-J. Breitkreutz, T.~Reguly, L.~Boucher, A.~Breitkreutz, and
  M.~Tyers (2006, Jan).
\newblock {BioGRID}: a general repository for interaction datasets.
\newblock {\em Nucleic Acids Res\/}~{\em 34\/}(Database issue), D535--D539.

\bibitem[\protect\citeauthoryear{Stelniec-Klotz, Legewie, Tchernitsa, Witzel,
  Klinger, Sers, Herzel, Blüthgen, and Schäfer}{Stelniec-Klotz
  et~al.}{2012}]{Stelniec-Klotz2012}
Stelniec-Klotz, I., S.~Legewie, O.~Tchernitsa, F.~Witzel, B.~Klinger, C.~Sers,
  H.~Herzel, N.~Blüthgen, and R.~Schäfer (2012).
\newblock Reverse engineering a hierarchical regulatory network downstream of
  oncogenic kras.
\newblock {\em Mol Syst Biol\/}~{\em 8}, 601.

\bibitem[\protect\citeauthoryear{Tresch and Markowetz}{Tresch and
  Markowetz}{2008}]{Tresch2008}
Tresch, A. and F.~Markowetz (2008).
\newblock Structure learning in nested effects models.
\newblock {\em Stat Appl Genet Mol Biol\/}~{\em 7}, Article9.

\bibitem[\protect\citeauthoryear{Wang, Castro, Mulder, and Markowetz}{Wang
  et~al.}{2012}]{Wang2012}
Wang, X., M.~A. Castro, K.~W. Mulder, and F.~Markowetz (2012).
\newblock Posterior association networks and functional modules inferred from
  rich phenotypes of gene perturbations.
\newblock {\em PLoS Comput Biol\/}~{\em 8\/}(6), e1002566.

\bibitem[\protect\citeauthoryear{Wang, Yuan, Hellmayr, Liu, and Markowetz}{Wang
  et~al.}{2013}]{Wang2013}
Wang, X., K.~Yuan, C.~Hellmayr, W.~Liu, and F.~Markowetz (2013).
\newblock Reconstructing evolving signaling networks by hidden markov nested
  effects models.
\newblock {\em Annals of Applied Statistics\/}~{\em acepted}, .

\bibitem[\protect\citeauthoryear{Werhli and Husmeier}{Werhli and
  Husmeier}{2007}]{Werhli2007}
Werhli, A.~V. and D.~Husmeier (2007).
\newblock Reconstructing gene regulatory networks with {B}ayesian networks by
  combining expression data with multiple sources of prior knowledge.
\newblock {\em Stat Appl Genet Mol Biol\/}~{\em 6}, Article15.

\bibitem[\protect\citeauthoryear{Workman, Mak, McCuine, Tagne, Agarwal, Ozier,
  Begley, Samson, and Ideker}{Workman et~al.}{2006}]{Workman2006}
Workman, C.~T., H.~C. Mak, S.~McCuine, J.-B. Tagne, M.~Agarwal, O.~Ozier, T.~J.
  Begley, L.~D. Samson, and T.~Ideker (2006, May).
\newblock A systems approach to mapping dna damage response pathways.
\newblock {\em Science\/}~{\em 312\/}(5776), 1054--1059.

\bibitem[\protect\citeauthoryear{Yates and Mukhopadhyay}{Yates and
  Mukhopadhyay}{2013}]{Yates2013}
Yates, P.~D. and N.~D. Mukhopadhyay (2013).
\newblock An inferential framework for biological network hypothesis tests.
\newblock {\em BMC Bioinformatics\/}~{\em 14}, 94.

\bibitem[\protect\citeauthoryear{Yeang, Mak, McCuine, Workman, Jaakkola, and
  Ideker}{Yeang et~al.}{2005}]{Yeang2005}
Yeang, C.-H., H.~C. Mak, S.~McCuine, C.~Workman, T.~Jaakkola, and T.~Ideker
  (2005).
\newblock Validation and refinement of gene-regulatory pathways on a network of
  physical interactions.
\newblock {\em Genome Biol\/}~{\em 6\/}(7), R62.

\bibitem[\protect\citeauthoryear{Yip, Alexander, Yan, and Gerstein}{Yip
  et~al.}{2010}]{Yip2010}
Yip, K.~Y., R.~P. Alexander, K.-K. Yan, and M.~Gerstein (2010).
\newblock Improved reconstruction of in silico gene regulatory networks by
  integrating knockout and perturbation data.
\newblock {\em PLoS One\/}~{\em 5\/}(1), e8121.

\end{thebibliography}

\end{document}